\documentstyle[12pt,twoside,fleqn,epsfig,espcrc1]{article}

\newcommand{\AmS}{{\protect\the\textfont2
  A\kern-.1667em\lower.5ex\hbox{M}\kern-.125emS}}
\newcommand{\be}{\begin{equation}}
\newcommand{\ee}{\end{equation}}
\newcommand{\bea}{\begin{eqnarray}}
\newcommand{\eea}{\end{eqnarray}}
\newdimen\rotdimen
\def\vspec#1{\special{ps:#1}}
\def\rotstart#1{\vspec{gsave currentpoint currentpoint translate
   #1 neg exch neg exch translate}}
\def\rotfinish{\vspec{currentpoint grestore moveto}}
\def\rotr#1{\rotdimen=\ht#1\advance\rotdimen by\dp#1%
   \hbox to\rotdimen{\hskip\ht#1\vbox to\wd#1{\rotstart{90 rotate}%
   \box#1\vss}\hss}\rotfinish}
\def\rotl#1{\rotdimen=\ht#1\advance\rotdimen by\dp#1%
   \hbox to\rotdimen{\vbox to\wd#1{\vskip\wd#1\rotstart{270 rotate}%
   \box#1\vss}\hss}\rotfinish}%
\def\rotu#1{\rotdimen=\ht#1\advance\rotdimen by\dp#1%
   \hbox to\wd#1{\hskip\wd#1\vbox to\rotdimen{\vskip\rotdimen
   \rotstart{-1 dup scale}\box#1\vss}\hss}\rotfinish}%
\def\rotf#1{\hbox to\wd#1{\hskip\wd#1\rotstart{-1 1 scale}%
   \box#1\hss}\rotfinish}%

\title{Relations between Electromagnetic 
Form Factors of Baryons}

\author{A. J. Buchmann\address{Institute for Theoretical Physics,
        University of T\"ubingen, Auf der Morgenstelle 14 \\
        D-72076 T\"ubingen, Germany}%
        \thanks{email:alfons.buchmann@uni-tuebingen.de}
        \thanks{Talk given at Intern. Symposium on Physics of Hadrons and 
Nuclei,
Tokyo, Japan, 14-17 Dec., 1998, published in Nucl. Phys. {\bf A670} (2000) 
174c-177c.} }

\begin{document}

%
%
\def\shiftleft#1{#1\llap{#1\hskip 0.04em}}
\def\shiftdown#1{#1\llap{\lower.04ex\hbox{#1}}}
\def\thick#1{\shiftdown{\shiftleft{#1}}}
\def\b#1{\thick{\hbox{$#1$}}}

\maketitle
\begin{abstract}
The inclusion of two-body exchange currents in the constituent quark model
leads to new relations between the electromagnetic properties of 
octet and decuplet baryons. In particular, the $N \to \Delta$ quadrupole 
transition form factor can be expressed in terms of the neutron
charge form factor.
\end{abstract}

\section{Neutron and $\Delta$ charge form factors} 

\begin{figure}[h]
$$\mbox{
\epsfxsize 10.5 true cm
\epsfysize 3.0 true cm
\setbox0= \vbox{
\hbox { \centerline{
\epsfbox{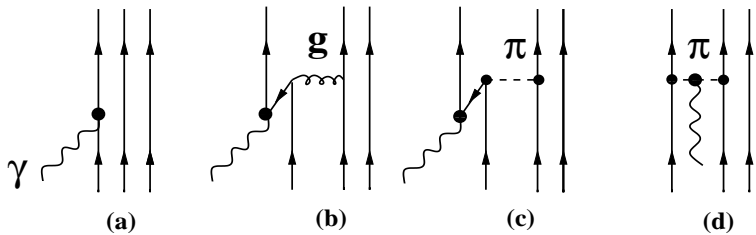}
}} 
} 
\box0
} $$
\vspace{-1.2cm}
\caption[Exchange currents]{Feynman diagrams of the four vector
current $J^{\mu}=(\rho, {\bf J})$:
photon ($\gamma$) coupling to (a) one-body current $J^{\mu}_{[1]}$, 
and to (b-d) two-body gluon and pion exchange currents $J^{\mu}_{[2]}$.
Diagrams (b-d) must be taken into account in order to satisfy the continuity
equation $q_{\mu}J^{\mu}=0$ for the electromagnetic current 
$J_{\mu}$. They represent the nonvalence (gluon and pion) degrees of 
freedom in the nucleon.}
\label{feynmec}
\end{figure}

In Ref. \cite{Buc91} we have shown that the Sachs charge form factor
$G_E^n(q^2)$ and charge radius 
$r_n^2=-6 (d/d{\bf q}^2) G_E^n({\bf q}^2) \mid_{{\bf q}^2=0}$  of the neutron 
are dominated by quark-antiquark pair exchange currents shown in 
Fig.\ref{feynmec}(b-c). The latter
describe the gluon and pion degrees of freedom, while the 
one-body currents in Fig.\ref{feynmec}(a) describe 
the valence quark degrees of freedom in the nucleon. 
The two-body exchange charge operator 
contains a spin-dependence of the schematic form 
$\rho_{[2]} \propto \b{\sigma}_i \cdot \b{\sigma}_j 
Y^0({\bf q}) -{\sqrt{6} \over 2} \, 
[\,[\b{\sigma}_i \times \b{\sigma}_j]^2 \times Y^2({\bf q})\, ]^0$. 
This gives different matrix elements for quark pairs
in spin 0 and spin 1 states. Because a down-down pair in the neutron 
is always in a spin 1 state, while an up-down pair can be in spin 0 
or spin 1 states, an asymmetry in the charge distribution of up and down 
quarks arises. Consequently, a nonvanishing
neutron charge form factor (see Fig.2) and radius is obtained 
\cite{Buc91}: 
\be
\label{rpmec}
r^2_{n}   =  -{b^2\over 3m_q} \biggl ( \delta_g(b)+ \delta_{\pi}(b) \biggr )
=-b^2 { M_{\Delta}-M_N \over M_N }. 
\ee
Here, $b$ is the quark core (matter) radius of the nucleon, $m_q$ is
the constituent quark mass, and the functions $\delta_g(b)$ and 
$\delta_{\pi}(b)$ are the gluon and pion contributions to the $N$-$\Delta$ 
mass splitting, which satisfy $M_{\Delta}-M_N=\delta_g(b)+\delta_{\pi}(b)$. 

\begin{figure}[t]
$$\hspace{0.2cm} \mbox{
\epsfxsize 9.0 true cm
\epsfysize 12.0 true cm
\setbox0= \vbox{
\hbox {
\epsfbox{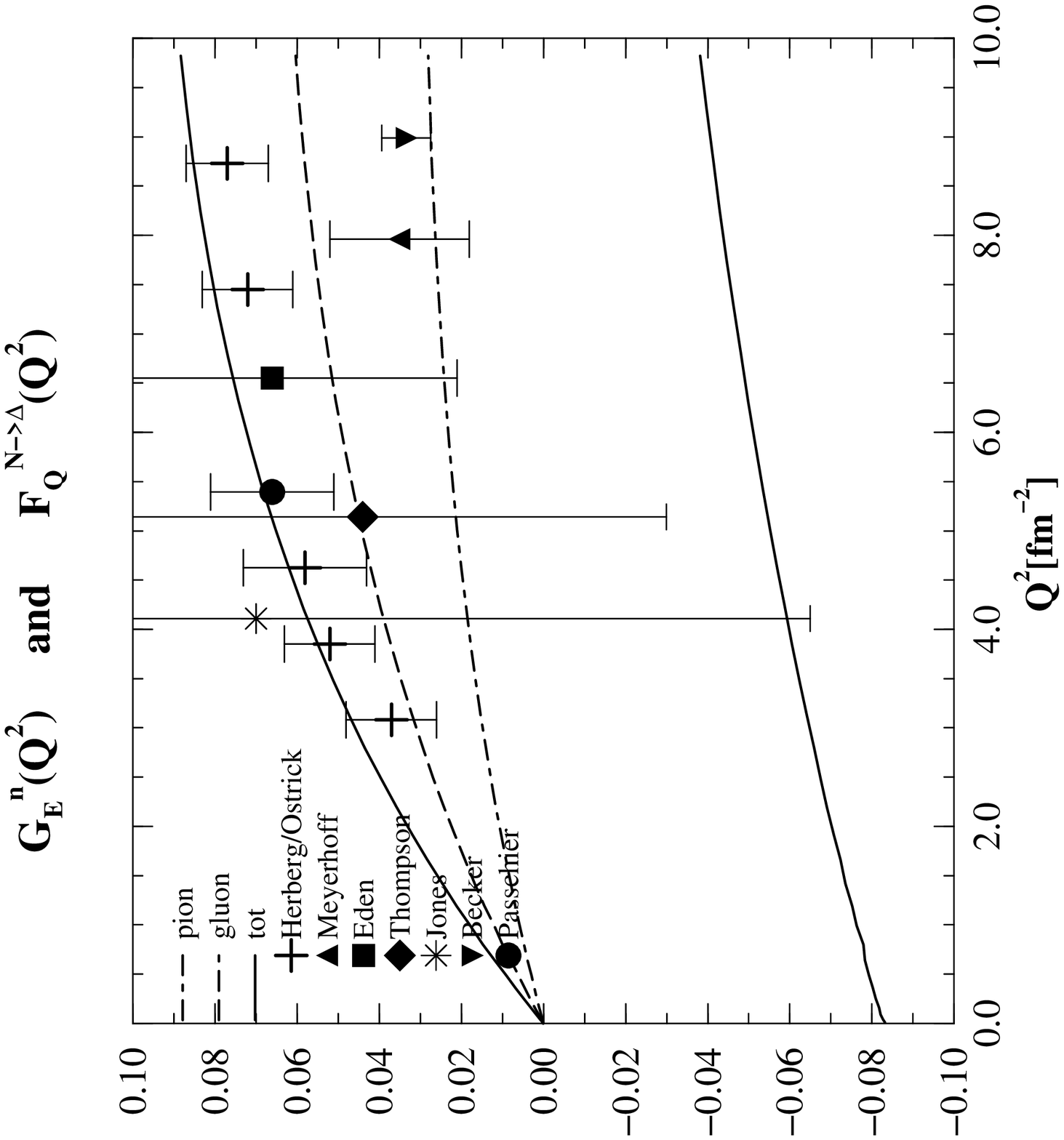}
} 
} 
\rotr0
} $$
\vspace{-1.8cm}
\caption[Interpretation]{
Neutron charge form factor $G_E^n(Q^2)$ and the
quadrupole transition form factor 
$F_Q^{p \to \Delta^+}(Q^2)$ as a function of four-momentum transfer 
$Q^2=-q^2$.
The crosses, triangles, circles are the recent data \cite{Her99}. 
The upper curve is a quark model calculation with exchange currents 
\cite{Buc91,Buc97b}. The gluon and pion contributions to 
$G^n_E(Q^2)$ are shown separately. 
The lower curve is $F_Q^{p \to \Delta^+}(Q^2)=-3\sqrt{2}\, \, 
G_E^n(Q^2)/Q^2$.}
\label{nchaff99}
\end{figure}

The spin-spin term in the charge operator is also responsible for the
following relation between the charge form factors of the nucleon and
$\Delta$: 
\be
\label{crdn}
G_E^{p}({\bf q}^2)-G_E^{\Delta^+}({\bf q}^2)= G_E^{n}({\bf q}^2), \quad
r^2_p-r^2_{\Delta^+} = r^2_n, 
\ee
where $r^2_{\Delta^+}$ is the charge radius
of the $\Delta^+$ and ${\bf q}$ is the three-momentum transfer of the photon.
An analogous result holds for the difference of neutron and $\Delta^0$
form factors.
The charge form factor of the $\Delta^0$ and the corresponding charge radius
are zero in the present model as it should be on general grounds \cite{Col85}.
From its derivation \cite{Buc97} it is evident that the charge radius 
relation is independent of whether gluon or pion exchange is the dominating 
residual interaction between constituent quarks. 
Using a general parametrization method, Dillon and Morpurgo \cite{Mor99} have  
recently shown that, if three-quark currents and strange quark loops  
are neglected, $r_p^2 -r_{\Delta^+}^2= r_n^2$ 
is a consequence of the symmetries and dynamics of QCD 
that is model-independently valid.
They have also shown that 
three-body currents slightly modify, but do not invalidate the general
relation between the proton, neutron, and $\Delta$ charge radii.

We repeat that Eq.(\ref{crdn}) is a
consequence of the spin-spin $\b{\sigma}_i \cdot \b{\sigma}_j/(m_i m_j)$
term in $\rho_{[2]}$, which leads to a $\Delta$ charge radius that is larger  
than the proton charge radius. This effect is of the same generality as,
and closely connected with the $N-\Delta$ mass splitting due to the 
spin-spin interaction in the Hamiltonian. The latter is repulsive 
in quark pairs with 
spin 1 and makes the $\Delta$ heavier than the nucleon.\footnote{Combining
Eq.(\ref{rpmec}) and Eq.(\ref{crdn}) yields 
$r^2_{\Delta^+}-r_p^2 =[b^2/(3 m_q)] (M_{\Delta^+}-M_p )$, i.e., 
a relation between the mass difference between $\Delta^+$ and proton
and a corresponding charge radius difference.}

\goodbreak 
\section{Electromagnetic $N \to \Delta$ transition form factors}
\nobreak

In the constituent quark model with exchange currents a connection
between the neutron charge form factor $G_E^n({\bf q}^2)$ \cite{Buc91}
and the $N\to\Delta$ quadrupole transition 
form factor $F_Q^{p \to \Delta^+}({\bf q}^2)$ \cite{Buc97b,Buc97} emerges:
\be
\label{qnff} 
F_Q^{p \to \Delta^+}({\bf q}^2)= -{3 \sqrt{2}\over {\bf q}^2}  \, 
G_E^n({\bf q}^2), \quad 
Q_{p \to \Delta} ={ r_n^2 \over \sqrt{2}}, \quad
r^{2}_{Q, \, p \to \Delta^+}= {11\over 20}\, b^2 +r^2_{\gamma q}.
\ee
The $N\to\Delta$ quadrupole transition form factor is a measure
of the intrinsic deformation of the nucleon and the $\Delta$.
The above results for the $N \to \Delta$ transition quadrupole moment, 
$Q_{p \to \Delta^+}$, and the  transition quadrupole 
radius\footnote{The term ${1 \over 4}b^2$ in Eq.(53) of 
Ref.\cite{Buc97b} should be replaced 
by ${11 \over 20} b^2$}, $r^2_{Q, \, p \to \Delta^+}$,
were derived before\cite{Buc97b,Buc97}. 
They are seen here to be the 0th and 1st moment of the 
more general relation between $G_E^n$ and $F_Q^{p \to \Delta^+}$,
plotted as the lower curve in Fig.\ref{nchaff99}.
The quark model with exchange currents explains $Q_{p \to \Delta^+}$
as a {\it double spin flip} of two quarks, with all valence quarks 
remaining in the dominant, spherically symmetric $L=0$ state. 
The double spin-flip comes from the tensor term in $\rho_{[2]}$.
The latter is closely related to the tensor term in the Hamiltonian.
The quark core ($D$ waves in the nucleon) also contributes  
to $Q_{p \to \Delta^+}$. This valence quark contribution 
amounts to about 20$\%$ (due to the smallness of the $D$ wave amplitudes)
of the double spin flip amplitude \cite{Buc97}.
We conclude that the collective gluon and pion degrees of freedom are mainly
responsible for the deformation of the $N$ and $\Delta$.

Due to the first relation in Eq.(\ref{qnff}), the quadrupole 
transition radius can also be expressed 
as the 2nd moment of $G_E^n({\bf q}^2)$, 
namely, $r^2_{Q,\, p \to \Delta^+}= ({18/r_n^2}) \, {(d/d{\bf q}^2)^2}
G_E^n({\bf q}^2)\bigl \vert_{{\bf q}^2=0}$.  
Because the quark core radius $b$ is fixed by Eq.(\ref{rpmec}), 
one could extract the charge radius of the light constituent 
quarks, $r^2_{\gamma q}$, from 
both the $G_E^n({\bf q}^2)$ data, and from the slope 
of $F_Q^{p\to\Delta^+}({\bf q}^2)$ 
at ${\bf q}^2=0$. Both determinations of $r^2_{\gamma q}$ 
should agree.

It is interesting that the additive quark model relation between the 
magnetic $N\to \Delta$  transition and the neutron magnetic moments 
$\mu_{p \to \Delta^+}=-\sqrt{2} \mu_n$ remains unchanged after including
the gauge-invariant two-body exchange currents of Fig.\ref{feynmec}(b-d); 
and that it continues to hold even at finite momentum transfers
\be
\label{magff} 
F_M^{p \to \Delta^+}({\bf q}^2)= -\sqrt{2} \, G_M^n({\bf q}^2), \qquad
\mu_{p\to \Delta^+}= -\sqrt{2} \, \mu_n, \qquad
r^2_{M, \, p \to \Delta^+}= -\sqrt{2} \, 
{ \mu_n \over \mu_{p \to \Delta^+}} \, r^2_{M, \, n}, 
\ee
where $r^2_{M, \, p \to \Delta^+}$ is the magnetic 
$N \to \Delta$ transition radius, and
$r^2_{M, \,  n}$ the magnetic radius of the neutron.
The transition magnetic moment predicted by Eq.(\ref{magff}) underestimates
the empirical value by 30$\%$. This discrepancy between theory and experiment
can presumably be explained by including spatial three-body 
currents in the theoretical description \cite{Mor89}. 

Combining Eq.({\ref{qnff}) and Eq.(\ref{magff}) we find that the 
ratio of the charge quadrupole and magnetic dipole 
$N\to \Delta$ transition form factors can be expressed 
in terms of the experimentally better known elastic neutron form factors
\be
\label{c2m1}
{F_Q^{p \to \Delta^+}({\bf q}^2) \over F_M^{p \to \Delta^+}({\bf q}^2)}=
{3 \over {\bf q}^2} \,  {G_E^{n}({\bf q}^2) \over G_M^{n}({\bf q}^2)},
\qquad 
{C2 \over M1} = {M_N \omega_{cm} \over 2 \, {\bf q}^2 }{ G_E^n({\bf q}^2)
\over G_M^n({\bf q}^2)},   
\ee
where $\omega_{cm}=258$ MeV is the center of mass energy
of the photon-nucleon system at the $\Delta$ resonance.
For example, this yields 
${C2\over M1} ({\bf q}^2\!\!=\!\!0)=- 1.04 \,\, {r_n^2/(2\, \mu_n)}=-0.030$ and
${C2 \over M1}({\bf q}^2\!\!=\!\!4.2\, {\rm fm}^{-2})=-0.042$. 
Sign and magnitude
of these theoretical predictions are in agreement with recent 
experimental data 
${C2 \over M1}({\bf q}^2\!\!=\!\!4.2\, {\rm fm}^{-2})_{exp}=-0.046(8)$ 
\cite{Bar99}.

\goodbreak 
\section{Relations between octet and decuplet hyperon charge radii}
\nobreak

Using the general parametrization method of Refs.\cite{Mor99,Mor89}, 
we find the following 
relations between octet and decuplet charge radii for strange hyperons:
\be
\label{newrel}
r_{\Sigma^{-}}^2-r_{\Sigma^{*-}}^2=r_{\Xi^{-}}^2-r_{\Xi^{*-}}^2=
r_n^2\left (x + x^2 \right ), \qquad 
r_{\Sigma^{+}}^2-r_{\Sigma^{*+}}^2=r_{\Xi^{0}}^2-r_{\Xi^{*0}}^2=
r_n^2\left (2x - x^2 \right ),
\ee
where $x=m_u/m_s$ is the ratio of nonstrange to strange quark masses.
Again, it is the $\b{\sigma}_i \cdot \b{\sigma}_j/(m_i m_j) $ term in the 
charge operator that leads to Eq.(\ref{newrel}).

In summary, by including two-body currents 
in the constituent quark model we have found hitherto unknown relations 
between the electromagnetic form factors of octet and decuplet baryons.
In particular, 
the $C2/M1$ ratio in the electromagnetic 
$N \to \Delta$ transition can be expressed in terms of 
the elastic form factors of the neutron.

\end{document}